\documentstyle[preprint,12pt,aps,psfig]{revtex}

\tighten
\begin{document}
\draft
\preprint{DAMTP/97/11, $\;\;\;$ IMPERIAL/TP/96-97/28}

\newcommand{\nc}{\newcommand}
\nc{\al}{\alpha}
\nc{\ga}{\gamma}
\nc{\de}{\delta}
\nc{\ep}{\epsilon}
\nc{\ze}{\zeta}
\nc{\et}{\eta}
\renewcommand{\th}{\theta}
\nc{\ka}{\kappa}
\nc{\la}{\lambda}
\nc{\rh}{\rho}
\nc{\si}{\sigma}
\nc{\ta}{\tau}
\nc{\up}{\upsilon}
\nc{\ph}{\phi}
\nc{\ch}{\chi}
\nc{\ps}{\psi}
\nc{\om}{\omega}
\nc{\Ga}{\Gamma}
\nc{\De}{\Delta}
\nc{\La}{\Lambda}
\nc{\Si}{\Sigma}
\nc{\Up}{\Upsilon}
\nc{\Ph}{\Phi}
\nc{\Ps}{\Psi}
\nc{\Om}{\Omega}
\nc{\ptl}{\partial}
\nc{\del}{\nabla}
\nc{\be}{\begin{eqnarray}}
\nc{\ee}{\end{eqnarray}}
\nc{\lambar}{\overline{\lambda}}


\title{Restoration of Isotropy for Spin Models} 

\author{
Dorje Brody$^*$\footnote{Electronic address: d.brody@damtp.cam.ac.uk}
and 
Adam Ritz$^{\dagger}$\footnote{Electronic address: a.ritz@ic.ac.uk}
} 
\address{* Department of Applied Mathematics and Theoretical 
Physics, \\ University of Cambridge, Silver Street, Cambridge CB3 9EW U.K.}
\address{$\dagger$ Theoretical Physics Group, Blackett Laboratory, \\
   Imperial College, South Kensington, London SW7 2BZ U.K.} 
\date{\today}

\maketitle

\vspace{0.3in}

\begin{abstract} 
Using real-space renormalisation techniques we analyse the Ising 
model on a Sierpi\'nski gasket with anisotropic microscopic 
couplings, and observe a restoration of isotropy on macroscopic 
scales. In particular, via use of a decimation 
procedure directly on the fractal lattice, we calculate explicitly 
the exponential anisotropy decay coefficients near the isotropic 
regime for both ferromagnetic and antiferromagnetic systems. The 
results suggest the universality of the phenomenon in lattice 
field theories on fractals.
\end{abstract}


\section{Introduction} 

Restoration of macroscopic isotropy in a system which has an 
underlying microscopic anisotropy is a phenomenon which, while 
absent in uniform media, appears almost universal in fractals. 
Recently, Barlow et al. \cite{barlow95} considered this phenomenon 
in the form of resistor networks. In particular, they considered 
a resistor network constructed on the Sierpi\'nski gasket with 
locally anisotropic resistance elements. Then, by successive use 
of star-triangle transformations they obtained recursion relations 
for the resistance elements from one length scale to another, and 
observed a restoration of isotropy at the macroscopic scale. The 
extensions to other fractal objects were also considered and the 
results suggest the universality of the phenomenon at least on 
exactly self--similar fractals. 
A question which naturally arises in this regard is whether the 
same mechanism is present for systems in statistical mechanics, 
or quantum field theories. \par

The statistical mechanics of spin systems on fractal lattices, on the 
other hand, has been a subject of considerable interest for which 
a number of models has been studied. In particular, for
exactly self--similar fractals with a finite degree of ramification, 
exact real-space renormalisation techniques can 
be pursued for simple spin models, and as a consequence the Ising spin 
model with isotropic couplings has been solved exactly on various 
fractals. See, for example, Gefen et. al. \cite{gefen80,gefen84}, 
Luscombe and Desai \cite{luscombe85a} or Stosi\'c et. al. 
\cite{stosic88}
for further details. The physical motivation behind these 
investigations is in part that the results obtained on fractals may 
have some relevance for real random systems, or crystals with 
defects. Also, it is interesting to note that this analysis 
indicates properties of spin systems in nonintegral dimensions 
which are not apparent in more formal analyses such as the 
$\ep$--expansion, and also provides the possibility for 
considering how this physical behaviour crosses over to
uniform media \cite{stosic88}. \par 

In the present Letter, we study the Ising model on a Sierpi\'nski 
gasket with locally anisotropic coupling configurations. As a 
comparison to the formulation presented in \cite{barlow95}, we first 
analyse this model with a successive use of star-triangle 
transformations and one-dimensional decimation, in order 
to study the level of anisotropy at larger length scales. The 
results obtained show enormously rapid recovery of isotropy 
at larger scales. However, unlike the case of resistance 
networks, the use of the star-triangle transformation in the 
present context is not necessarily advantageous in the sense 
that frustrated spin systems cannot be studied within this 
methodology. Nevertheless, as we noted above, in the case of the 
Ising model, exact real-space renormalisation transformations can 
also be performed in the frustrated system (see for example
Stinchcombe \cite{stinchcombe89} and Grillon \& Brady Moreira
\cite{grillon89}), and as a consequence we are able 
to study the large scale structure for both ferromagnetic and 
antiferromagnetic Ising models. In both cases, we observe recovery 
of isotropy at an exponential rate. 
More specifically, the technique used allows us to rigorously 
prove the convergence of the system to isotropy and to investigate 
the rate at which homogeneity is restored. In particular in the near 
isotropic regime the scaling to isotropy is exponential with a 
coefficient given by $\rh=1\pm\tanh2$ with the upper(lower) sign 
corresponding to the ferro(antiferro)-magnetic system.\par 

\section{Results from star-triangle transformations} 

The model is constructed as follows. We embed the Sierpi\'nski 
gasket in two Euclidean dimensions and place Ising spins 
$\si_i=\pm 1$ at each vertex. Geometrically the Ising model is 
then effectively carried by a space of Hausdorff dimension 
$d = \ln3/\ln2 = 1.585$ with a finite degree of ramification 
$R_{max} = 4$.We consider the standard nearest neighbour 
Hamiltonian given by 
\begin{eqnarray}
{\cal H}\ =\ - \sum_{(ij)}K_{ij}\si_i\si_j\ , 
\end{eqnarray}
where $K_{ij}=J_{ij}/kT$ are anisotropic ferromagnetic couplings 
which depend upon the orientation of the spins $\si_i$ and $\si_j$, 
and the summation is taken over all the nearest neighbour sites on the 
fractal lattice. 
\begin{figure}
\label{stri}
 \centerline{%
   \psfig{file=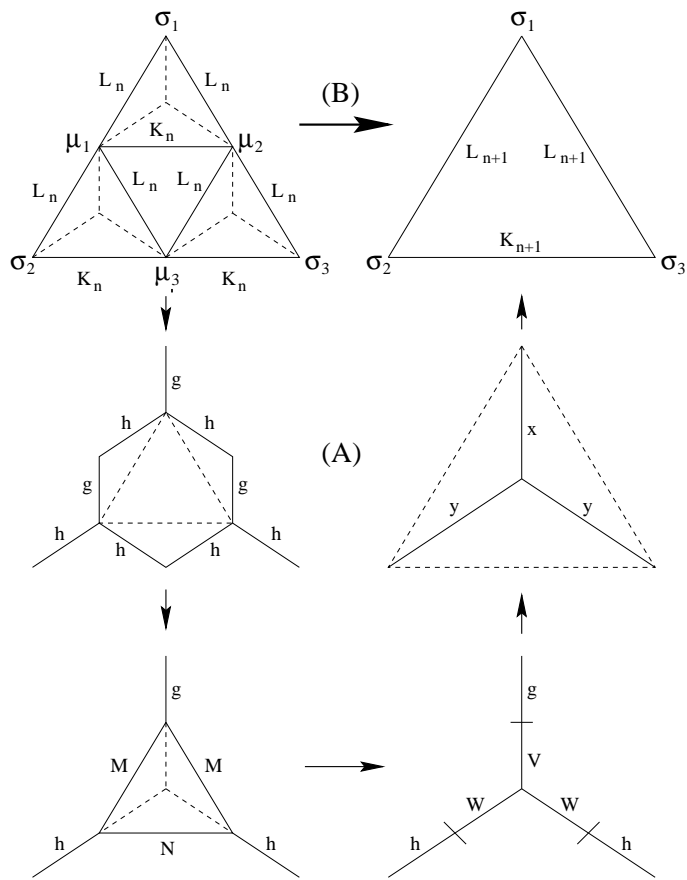,width=9cm,angle=270}%
   }
 \vspace{0.3in}
 \caption{A stage in the renormalisation group procedure for the Ising model 
          on the SG is indicated. The interior spins $\mu_1,\mu_2,\mu_3$ 
          are integrated out resulting in course grained couplings 
          $L_{n+1}$ and $K_{n+1}$. This is achieved in (A): via a 
          combination of star-triangle transformations and 1D decimation,
          the dashed lines indicating the transformation used at each stage 
          (the various couplings being defined in the text); while in (B):
          direct decimation on the fractal lattice is used.}
\end{figure}
\noindent Motivated by the technique used in \cite{barlow95} 
we  can construct a real-space renormalisation group procedure 
as indicated in Fig.~1 [part (A)] where after the n$^{th}$ 
iteration of the process we denote the horizontal coupling 
by $K_n^{hor}=K_n$, and the two `vertical' couplings as 
$K_n^{vert1}=K_n^{vert2}=L_n=r_nK_n$ so that $r_n$ represents the 
degree of anisotropy after the n$^{th}$ iteration. \par 

The process involves the use of an alternate sequence of star-triangle 
transformations and one-dimensional decimation so that if we define 
the following functions
\begin{eqnarray}
\left\{ \begin{array}{l} 
F_{1}(\al,\beta)\ =\ \frac{1}{2}{\rm arccosh}\left[e^{2\al} 
\cosh2\beta\right] \\ 
F_{2}(\al,\beta)\ =\ {\rm arctanh}\left[\frac{e^{2\al}\sinh2\beta} 
{\sinh(2F_1(\al,\beta))}\right] \\ 
F_{3}(\al,\beta)\ =\ \frac{1}{2}\ln\left[\frac{\al+\beta}{\al - 
\beta}\right] \\
F_{4}(\al,\beta)\ =\ \frac{1}{4}\ln\left[\frac{2\al+\beta} 
{2\al-\beta}\right] \\
F_{5}(\al,\beta)\ =\ \frac{1}{2}\ln\left[\cosh^22\beta 
-\sinh^22\beta\tanh^2\al\right]\ , 
\end{array} \right. 
\end{eqnarray}
then the couplings at each stage of the process indicated in 
Fig.~1 are given by
\begin{eqnarray} 
\left\{ \begin{array}{ll} 
g\ =\ F_1(K_n,L_n),     &  \    h\ =\ F_2(K_n,L_n) \\ 
M\ =\ F_3(g,h),         &  \    N\ =\ F_3(h,h) \\ 
W\ =\ F_1(N,M),         &  \    V\ =\ F_2(N,M) \\ 
x\ =\ F_3(g,V),         &  \    y\ =\ F_3(h,W) \\ 
K_{n+1}\ =\ F_4(x,y),   &  \    L_{n+1}\ =\ F_5(x,y).
\end{array} \right. 
\end{eqnarray}
This procedure may be iterated numerically, and with the initial 
conditions $K_0=0.01$, $L_0=1$, i.e. an initial microscopic
anisotropy level of $r_0=100$, the results we obtain are shown in Fig.~2. 
The simulation indicates a rapid 
restoration of isotropy although the actual structure of the 
anisotropy oscillates after each iteration, due to the fact that 
an odd number of star-triangle transformations are used during each 
iteration process, as indicated in Fig.~1. 

One can easily observe, from the form of the functions $F_1$ and 
$F_2$, that a `cut' along the negative real axis of the coupling 
space prevents the consideration of negative couplings and 
therefore a frustrated system cannot be treated via this procedure. 
In other words, if the initial couplings for each triangle are 
all negative (antiferromagnetic), then the 
resulting couplings on the `star' become pure imaginary.
\begin{figure}
\label{st}
 \centerline{%
   \psfig{file=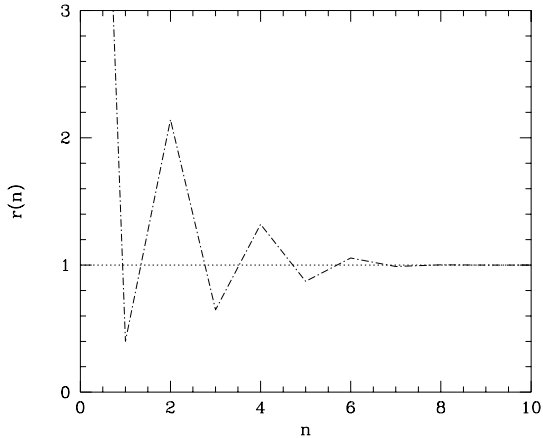,width=7.5cm,angle=0}%
   }
 \vspace{0.3in}
 \caption{The level of anisotropy $r_n$ plotted against the order of
          the coarse--graining iteration of type (A). Starting from 
initial ratio $r_{0}=100$, we recover $r\sim O(1)$ after only three 
iterations. }
\end{figure}
\noindent Physically, 
this reflects the fact that, while on the triangular lattice there 
are frustrations associated with each plaquette, these frustrations 
cannot be realised for any real-valued couplings on the `star', or 
equivalently, a hexagonal lattice. \par 

\section{Ferromagnetic case} 

In order to allow us to analyse more general situations, and also 
consider the possibility of obtaining analytic results for the 
asymptotic behaviour, we now consider a different method, based on 
the direct use of real-space renormalisation transformations. As 
mentioned above, due to the finite degree of ramification, exact 
renormalisation group analysis is easily accomplished analytically 
via decimation, as reported in 
\cite{gefen80,gefen84,luscombe85a}. 
A course graining transformation of this form is 
represented in part (B) of Fig.~1.

Consideration of the partition function allows the recursion relations
for the couplings $L$ and $K$ to be determined as follows,
\begin{eqnarray} 
e^{4K_{n+1}} =  \frac{2\cosh2K_{n} + \cosh 2L_n 
        +e^{4K_n}\cosh 6L_n}{2\cosh^2 2L_n(1+e^{4K_n}\cosh 2L_n)}
          \label{kn}
\end{eqnarray} 
\begin{eqnarray} 
e^{4L_{n+1}} = \frac{2\cosh2K_{n} + \cosh 2L_n
        +e^{4K_n}\cosh 6L_n}{2\cosh2K_{n} (1+e^{4K_n}\cosh 2L_n)}
           \label{ln}, 
\end{eqnarray}
which, in the isotropic limit $L_{n}=K_{n}$, clearly reduce to the 
result of Gefen et al. \cite{gefen80,gefen84},
\begin{eqnarray}
 e^{4K_{n+1}}\ =\ \frac{e^{8K_n}-e^{4K_n}+4}{e^{4K_n}+3}.
\end{eqnarray}
After $n$ iterations of the renormalisation group transformation 
the level of anisotropy in the interaction is again characterised by 
$r_n=L_n/K_n$. Since we are concerned only with this ratio of the 
couplings we can rescale both the horizontal and `vertical' 
couplings after each iteration by an arbitrary factor. A convenient 
choice of this factor proves to be $1/K_n$, and as a consequence the 
couplings at each order now take on the values $K_n=1$ and $L_n=r_n$. 
Clearly this rescaling has no effect on the observable behaviour of 
the system with regard to changes in the level of anisotropy. However, 
the benefit we gain by using this technique is that by taking the 
ratio of (\ref{kn}) and (\ref{ln}) we can construct a recursion 
relation purely for the level of anisotropy in the system given by
\begin{eqnarray}
 \label{recrn}
 r_{n+1} & = & 1+\frac{1}{2}\ln\left(\frac{\cosh 2r_n}{\cosh 2}\right).
\end{eqnarray}
Before we analyse this relation we should first consider the 
possible initial conditions. i.e., the initial choices one
can make for $r_0$. Ignoring the trivial isotropic
case there are clearly two; either $r_0>1$ or $r_0<1$. With reference
to Fig.1 we see that these conditions are not equivalent. The first
corresponds physically to the situation where couplings in two
of the directions on the lattice are larger than the third, while
the latter case corresponds to the situation where two of the 
couplings are weaker than the third. To avoid confusion we shall
always differentiate these particular initial conditions and
refer to each by the corresponding values of $r_0$.

The structure of the relation (\ref{recrn}) indicates that
it possesses a two element fixed point set $\{1,\infty\}$, with
the first point $r^*=1$, corresponding to isotropy. In order to 
determine which of these fixed points is stable we analyse the 
recursion relation for each of the possible initial conditions. 
Making use of the trivial inequality, $(a+b)/(c+d)<a/c$ for 
any $a,b,c,d \in {\bf R}_+$, we find 
\begin{eqnarray}
 r_{n+1}&<&1+\frac{1}{2}\ln\frac{e^{2r_n}}{e^2}=r_n, \;\;\;\;\;\;\;\;\;\;\;
               {\rm for }  \;\;\;r_n>1, \\
 r_{n+1}&>&1+\frac{1}{2}\ln\frac{e^{2r_n}}{e^2}=r_n, \;\;\;\;\;\;\;\;\;\;\;
               {\rm for } \;\;\;r_n<1.
\end{eqnarray}
\begin{figure}
\label{rn}
 \centerline{%
   \psfig{file=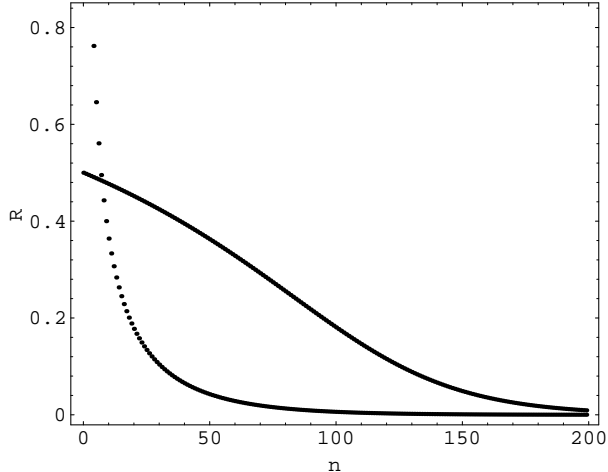,width=8cm,angle=0}%
   }
 \caption{The deviation from isotropy $R_n$ of the interaction plotted against
          the the renormalisation group iteration $n$ in the
          ferromagnetic case. The upper plot
          uses the initial condition $r_0=10$ while the lower curve
          uses $r_0=0.1$.} 
\end{figure}
\noindent Thus the deviation from isotropy at each order 
$\delta_n\equiv|r_n-1|$ satisfies the inequality 
$\delta_{n+1}<\delta_n$ unless $r_n=1$ and thus 
$r^*=1$ is the only stable fixed point of the recursion
relation and, as a consequence, restoration of isotropy 
on macroscopic scales is assured.

The actual behaviour of the system as one carries out the
course--graining transformations may be observed by iterating
(\ref{recrn}) numerically. Specifically, in Fig.~3 we present
the scaled deviation from isotropy 
\begin{equation} 
R_n\ =\ \left| 1-\frac{1}{r_{n}}\right| 
\end{equation} 
for examples of the possible initial
microscopic conditions $r_0=10$ and $r_0=0.1$. It is clear that
the presence of the unstable fixed point at $r^*=\infty$ will only
affect the restoration of isotropy in the first of these cases.
With the initial condition $r_0=0.1$, recovery of isotropy is
smooth and quite rapid, although approximately an order of
magnitude slower than the corresponding result using star--triangle 
transformations (see Section 2). For later reference it is useful
to characterise this regime as having `positive curvature'
(i.e., $r''(n)>0$ where we symbolically treat $r_n$ as a smooth
function). In contrast to the case $r_0=0.1$, with the initial
condition $r_0=10$ restoration of isotropy is quite
slow for several iterations and appears to be significantly 
affected by the fixed point at $r^{*}=\infty$. We characterise 
this regime as having `negative curvature' 
(symbolically $r''(n)<0$). We observe from Fig.~3 that after
sufficient iterations there is a crossover from this negative 
curvature regime to the positive curvature regime, corresponding 
to dominance of the $r^*=1$ fixed point, at approximately 
$r_n\approx 1.3$. As a consequence we can conclude that the 
negative curvature regime is only present if the initial 
configuration is such that $r_0>1.3$.

As we have mentioned above, the phenomenon of macroscopic 
restoration of isotropy appears to be an almost universal 
phenomenon in fractal media. In order to test this conjecture 
in a more quantitative manner it would be helpful to obtain 
a measure of the scaling near the isotropic fixed
point $r^*=1$. A numerical study of the recursion relation (\ref{recrn})
in this regime, for an arbitrary choice of initial conditions, indicates
that the rate of restoration of isotropy appears to be approximately
exponential. 
That this is indeed the case may be verified analytically as follows.
If we treat $r_n$ as a function, $r=r(n)$, 
we have in general $r(n+dn) = r(n)+r'(n)dn$ where the derivative 
is taken with respect to $n$. In the regime where $r_n\approx 1$ the 
variation in $r$ between iterations is very small and thus we may 
approximate the recursion relation (\ref{recrn}) by the truncated 
Taylor series, i.e., $r(n+1)\approx r(n)+r'(n)$. As a consequence, 
we obtain the differential equation
\begin{eqnarray}
 \label{diffr}
 \frac{{\rm d}r}{{\rm d}n} + r -\ln\frac{\cosh 2r}{\cosh 2}-1 & = & 0,
\end{eqnarray}
which, for arbitrary initial conditions, is valid in the regime 
near isotropy where $r$ is only weakly dependent on $n$. Recalling 
the definition of $\de=|r-1|$, in the limit $\de\rightarrow 0$ where 
the differential equation is valid, we may expand 
the nonlinear term in (\ref{diffr}) to obtain the following
simple differential equation for $\de$,
\begin{eqnarray}
 \frac{{\rm d}\de}{{\rm d}n} + (1+\tanh2)\de & = & 0,
\end{eqnarray}
with the solution given by 
\begin{eqnarray}
 \label{scale}
 \de(n) & = & \exp(-\rh n),
\end{eqnarray}
where the decay coefficient for the level of anisotropy
near the isotropic regime is given by $\rh=1+\tanh2\approx1.96$.

\section{Antiferromagnetic case} 

A significant advantage of the use of direct decimation on the
fractal lattice is that, in contrast to the results obtained
using star--triangle transformations, there are no `cuts' in the 
recursion relations (\ref{kn}) and (\ref{ln}) for negative values 
of the couplings. This allows us to adapt the technique to consider 
a frustrated system with negative couplings $K_0$ and $L_0$.

By rescaling the couplings after each renormalisation group 
iteration, as discussed in Section 3, we can derive a recursion 
relation for the level of anisotropy in a manner analogous to that 
described for the ferromagnetic case, obtaining
\begin{eqnarray}
 \label{recrnf}
 r_{n+1} & = & 1-\frac{1}{2}\ln\left(\frac{\cosh 2r_n}{\cosh 2}\right).
\end{eqnarray}
The possible initial conditions for this relation again fall into 
the same two classes, $r_0>1$ or $r_0<1$, which were relevant in 
the previous discussion. However, for this model the fixed point 
set for the recursion relation (\ref{recrnf}) reduces to a single 
point $\{1\}$ and from this fact alone one might expect that the 
two {\it a priori} different initial conditions will not lead to 
significantly different behaviour of the system under 
renormalisation. In other words, the existence of only one fixed 
point should lead to a single form of scaling behaviour.

Before we test this conjecture numerically we can verify the 
stability of the isotropic fixed point via arguments similar to 
those presented in Section 3. Treating the possible initial 
conditions separately we obtain 
\begin{eqnarray}
 r_{n+1}&>&1-\frac{1}{2}\ln\frac{e^{2r_n}}{e^2}=2-r_n, \;\;\;\;\;\;\;\;\;
               {\rm for }  \;\;\;r_n>1, \\
 r_{n+1}&<&1-\frac{1}{2}\ln\frac{e^{2r_n}}{e^2}=2-r_n, \;\;\;\;\;\;\;\;\;
               {\rm for }  \;\;\;r_n<1.
\end{eqnarray}
Thus the behaviour of $r_n$ in the frustrated case is oscillatory
about $r=1$ between iterations, which clearly accounts for the
loss of the fixed point at $\infty$ in this case. From these
relations one may again derive the inequality $\delta_{n+1}<\delta_n$ 
unless $r_n=1$ and thus $r^*=1$ is again a stable fixed point 
of the recursion relation and the system will tend to isotropy
on macroscopic scales. 
It is of interest to note that this is similar to the behaviour 
observed using star--triangle transformations in the ferromagnetic 
system. 

As anticipated, numerical iteration of the recursion relation 
(\ref{recrnf}) indicates that for the frustrated case there is no
qualitative difference in the behaviour of the deviation $R_n$
from isotropy due to a different choice of initial conditions. 
Specifically, in Fig.~4 the scaled deviation $R_n$ is plotted 
for the two initial configurations $r_0=0.1$ and $r_0=10$
indicating explicitly that the behaviour is qualitatively
the same in each case. This is again a consequence of the loss of 
the second fixed point at $r=\infty$.
\begin{figure}
\label{rnf}
 \centerline{%
   \psfig{file=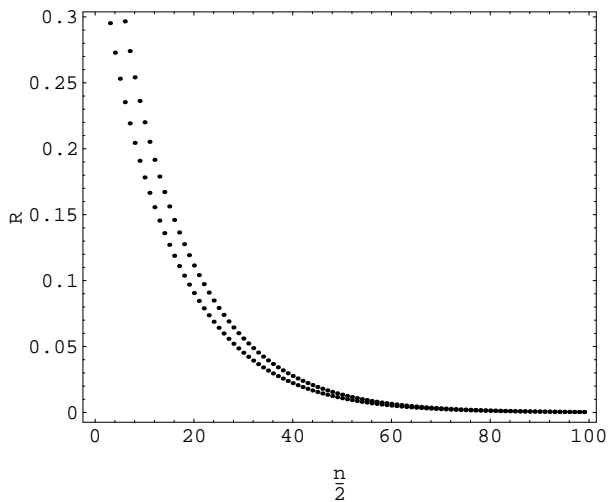,width=8cm,angle=0}%
   }
 \caption{The deviation from isotropy $R_n$ of the 
          frustrated interaction plotted against
          the the renormalisation group iteration $n$. The upper
          plot uses the initial condition $r_0=10$ while the lower curve
          uses $r_0=0.1$.} 
\end{figure}
\noindent Note that in this case, due 
to the fact that the oscillatory behaviour is not quite symmetric 
about $r=1$, we plot every second iteration for clarity.

The physical consequence of the loss of the second fixed point at 
infinity can easily be observed from Figures 3 and 4. In the 
situation where at microscopic scales we have $r_0>>1$, or in 
physical terms where the interaction of the Ising spins in one 
direction on the lattice is much weaker than than the other two, 
the restoration of isotropy in the frustrated system is 
significantly more rapid than the standard ferromagnetic system. 
Thus in this case the frustration appears to relieve the anisotropy 
on macroscopic scales.\par 

However, if we consider the regime near isotropy we obtain a
rather different conclusion. In particular if we analyse
the scaling of the `function' $r(n)$ in a manner analogous
to that presented in Section 3 we obtain a similar
differential equation to (\ref{diffr}), that is, 
\begin{eqnarray}
 \label{diffrf}
 \frac{{\rm d}r}{{\rm d}n} + r +\ln\frac{\cosh 2r}{\cosh 2}-1 & = & 
 0. 
\end{eqnarray} 
The deviation $\de_f=|r-1|$ in this case also satisfies the scaling 
relation (\ref{scale}), with a smaller decay coefficient 
$\rh=1-\tanh2\approx0.036$. Thus it appears
that in this regime the frustration inhibits homogenisation. As a 
consequence, one can envisage that at least in the situation described by 
an initial condition $r_0>>1$ the presence of a frustration 
has a markedly different effect on the level of isotropy
at different length scales when compared to the ferromagnetic
system. \par 

\section{Discussion} 

In conclusion, we have examined the phenomenon of restoration of 
macroscopic isotropy for a simple interacting system, a microscopically 
anisotropic Ising model on a finitely ramified fractal, the 
Sierpi\'nski gasket. This is an interesting example of a phenomenon
that is absent in uniform media. \par 

The results for the Ising spin model shows an exponential recovery 
of isotropy at large scales, for both ordered (ferromagnetic) and 
disordered (antiferromagnetic) systems. In particular, it is 
interesting to note that for disordered systems, the phenomenon is 
independent of the initial conditions, while for ordered systems the 
behaviour depends sensitively on the initial conditions. Furthermore, 
we have also observed that, in the strongly frustrated regime, the 
restoration of isotropy is considerably slower than the corresponding 
unfrustrated cases. \par 
 
The above results are also suggestive that further studies of 
models on spaces which are explicit geometric objects of 
nonintegral Hausdorff dimensions could uncover other subtle 
characteristics which may be of relevance to various physical 
systems. It could also be conjectured that the scaling behaviour 
which was analytically determined near isotropy may in fact be 
universal in such magnetic systems. \par 

The authors express their gratitude to B. Meister and 
N. Rivier for stimulating discussions. The financial support 
of D.B. by PPARC and A.R by the Commonwealth Scholarship 
Commission and the British Council is gratefully acknowledged. 

\bibliographystyle{prsty}

\end{document}